\documentclass[preprint,showpacs,preprintnumbers,amsmath,amssymb]{revtex4}

\def\mp{\textbf{\emph{p}}}
\usepackage{amssymb}
\usepackage{amsmath}
\usepackage{float}
\usepackage{graphicx}
\usepackage{subfigure}
\usepackage{color}

\begin{document}

\title{Normalized multi-pion Hanbury Brown-Twiss correlation functions of
pion-emitting sources with Bose-Einstein condensation}

\author{Ghulam Bary, Peng Ru, Wei-Ning Zhang\footnote{wnzhang@dlut.edu.cn}}
\affiliation{School of Physics, Dalian University of Technology, Dalian, Liaoning
116024, China}


\begin{abstract}
Recently, the ALICE collaboration analyzed the three- and four-pion Hanbury Brown-Twiss (HBT) correlations in Pb-Pb collisions at the Large Hadron Collider
(LHC). The measured suppressions of three- and four-pion correlations may
originate from a substantial coherence of the particle-emitting sources.
In this work we investigate the normalized three- and four-pion HBT correlation
functions for evolving pion gas (EPG) sources with Bose-Einstein condensation.
We find that the intercepts of the normalized correlation functions at zero
relative momentum are sensitive to source condensation and particle momentum.
The normalized correlation functions in low average-momentum regions of three
and four pions decrease with decreasing temperature and increasing
particle number of the source, indicating a dependence of the normalized
correlation functions on source condensation. However, this dependence becomes
weak in an intermediate average-momentum region because particles with high
momenta are likely emitted from excited states incoherently in the EPG model,
even if the source has a considerable condensation fraction. For a wide momentum
range, the normalized correlation functions for low source temperatures are
enhanced at larger relative momenta because of a rapid increase of two-pion
chaoticity parameter with increasing particle momentum. We hope the significant
enhancement of the normalized four-pion correlation function at high relative
momentum will be identified through future analyses of experimental data. 
\end{abstract}
\maketitle

\section{Introduction}
Two-pion Hanbury Brown-Twiss (HBT) interferometry is widely used to extract the
space-time structure of pion-emitting sources produced in high-energy heavy-ion
collisions \cite{Gyu79,Wongbook,Wie99,Wei00,Csorgo02,Lisa05}. One widely used
parameter in analyses of two-pion HBT interferometry is the chaoticity parameter,
$\lambda$, which is introduced by assuming a contribution of coherent particle
emission. The chaoticity parameter is also related to many other effects
in high-energy heavy-ion collisions, such as particle misidentification, final-state
Coulomb interaction, long-lived resonance decay, pion laser emission, and so on
\cite{Gyu79,Wongbook,Wie99,Wei00,Csorgo02,Lisa05,Pratt-PLB93,CsorgoZimanyi97}.

As an extension of two-pion interferometry, multi-pion interferometry has
been used in high-energy heavy-ion collisions \cite{{Wei00,Csorgo02,Pratt-PLB93,
CsorgoZimanyi97,Liu86,Zaj87,Biy90,And91-93,Zha93-00,ChaGaoZha95,HeiZhaSug,Nak99-00,
NA44,WA98,STAR-PRL03,Csa06,MorMurNak06,ALICE-PRC14,Gangadharan15,ALICE-PRC16,
BaryRuZhang-JPG18}}. The multi-pion correlation (MPC) analyses not only give an
alternative way to test the physics obtained by two-pion interferometry, but also
provide additional information of the particle-emitting sources. For example, the
triplet identical pion correlation includes the phase of source function and its
effect is important for asymmetric particle-emitting sources. More important, MPCs
are more sensitive to the source coherence compared to two-pion correlation.
In the heavy-ion collisions at the LHC energy, the identical pion multiplicity can
reach several thousand. The high pion multiplicity and the technical development
of MPC analysis \cite{Gangadharan15} open the door to accurately measure the MPCs
in experiment. Recently, the ALICE collaboration at the LHC find that there is a
significant suppression of MPCs in Pb-Pb collisions, and this suppression does
not be observed in the $pp$ and $p-$Pb collisions \cite{ALICE-PRC16}.
It may indicate that the suppression is a kind of medium effect of many particles.

In our previous work \cite{BaryRuZhang-JPG18}, we investigated the three- and
four-pion HBT correlation functions for heavy-ion collisions at the LHC
\cite{ALICE-PRC16}, based on an evolving pion gas (EPG) model with Bose-Einstein
condensation \cite{LiuRuZhangWong-JPG14}. Our model results of MPC functions were
consistent with experimental data and indicated a source condensation fraction
between 16\% and 47\% \cite{BaryRuZhang-JPG18}. Pion condensation may also enhance
the pion-transverse-momentum spectrum in low transverse-momentum region in heavy-ion
collisions at the LHC \cite{Begun14-15}.  However, to determine the source
condensation fraction with the HBT technique, one has to remove the other effects
on chaoticity parameters, especially the effect of long-lived resonance decay.

In Ref. \cite{HeiZhaSug}, Heinz, Zhang, and Sugarbaker proposed the
normalized three-pion correlation function $r_3$, which can be used to determine
the degree of source coherence without contamination from resonance decays.
The function $r_3$ has been used to analyze experimental data for heavy-ion
collisions at the CERN-SPS \cite{NA44,WA98}, RHIC \cite{STAR-PRL03}, and LHC
\cite{ALICE-PRC14}.
In this article, we investigate the normalized three-pion and four-pion correlation
functions, $r_3$ and $r_4$, in the EPG model for heavy-ion collisions at the LHC.
The results show that the normalized MPC functions in low average-transverse-momentum
region are sensitive to EPG source condensation.  The increase of the normalized
MPC functions at high relative momenta reflects the particle-correlation characteristic
in the EPG model, that the correlations decrease rapidly with increasing particle
momentum.

This article consists of four sections.  We present some basic MPC formulas and study
the intercepts of normalized MPC functions in the EPG model in section 2.  In section
3, we show and discuss the results of the normalized three- and four-pion correlation
functions in the EPG model.  Finally, we give a summary and discussions in section 4.

\section{Intercepts of normalized MPC functions in the EPG model}
By the definitions of HBT correlation functions with density matrices, the two-, three-,
and four-pion correlation functions can be written as \cite{BaryRuZhang-JPG18}
\begin{equation}
\label{Cp1p2}
C_2(\mp_1,\mp_2)=1+R_2(\mp_1,\mp_2),
\end{equation}
\begin{equation}
\label{Cp1p2p3}
C_3(\mp_1,\mp_2,\mp_3)=1+R_2(\mp_1,\mp_2)+R_2(\mp_1,\mp_3)+R_2(\mp_2,\mp_3)+R_3(\mp_1,
\mp_2,\mp_3),
\end{equation}
\begin{eqnarray}
\label{Cp1p2p3p4}
&&C_4(\mp_1,\mp_2,\mp_3,\mp_4)=1+R_2(\mp_1,\mp_2)+R_2(\mp_1,\mp_3)+R_2(\mp_1,
\mp_4)+R_2(\mp_2,\mp_3)\nonumber\\
&&\hspace*{33mm}+R_2(\mp_2,\mp_4)+R_2(\mp_3,\mp_4)+R_3(\mp_1,\mp_2,\mp_3)+R_3(\mp_1,
\mp_2,\mp_4)\nonumber\\
&&\hspace*{33mm}+R_3(\mp_1,\mp_3,\mp_4)+R_3(\mp_2,\mp_3,\mp_4)+R_2(\mp_1,\mp_2)
R_2(\mp_3,\mp_4)\nonumber\\
&&\hspace*{33mm}+R_2(\mp_1,\mp_3)R_2(\mp_2,\mp_4)+R_2(\mp_1,\mp_4)R_2(\mp_2,\mp_3)
\nonumber\\
&&\hspace*{33mm}+R_4(\mp_1,\mp_2,\mp_3,\mp_4)\!+\!R_4(\mp_1,\mp_2,\mp_4,\mp_3)\!
+\!R_4(\mp_1,\mp_3,\mp_2,\mp_4).
\end{eqnarray}
Here, $R_2(\mp_i,\mp_j)$, $[R_2(\mp_i,\mp_j)R_2(\mp_k,\mp_l)]$, $R_3(\mp_i,\mp_j,
\mp_k)$, and $R_4(\mp_i,\mp_j,\mp_k,\mp_l)$ denote the correlation of a single pion
pair, correlation of a double pion pair, pure pion-triplet interference or true
three-pion correlator \cite{Liu86,HeiZhaSug}, and pure pion-quadruplet interference,
respectively.

The particle-emitting source in the EPG model \cite{LiuRuZhangWong-JPG14} is a
quasi-static identical-pion gas trapped within a mean field with harmonic oscillator
potential \cite{WongZhang-PRC07} $\sim\!(\hbar\omega r^2/a^2)$, where $a=\sqrt{\hbar/
m\omega}$ is the characteristic length of the harmonic oscillator.  The harmonic
oscillator potential has been used to study Bose-Einstein condensation in atomic
physics \cite{Anderson-SCI95,NarGla-PRA99,Viana-PRA06}.  Its advantage here is that
the pion gas system can be analytically solved in nonrelativistic cases \cite{WongZhang-PRC07}, although the particle motion is relativistic in our model
calculations \cite{LiuRuZhangWong-JPG14,BaryRuZhang-JPG18}.
In the EPG model, the source evolution is assumed to be an adiabatic expansion
satisfying $TV^{\gamma-1}=$ constant at each state of evolution, which is an
approximation for the case that the system relaxation time is shorter than the source
evolution time.  Here, $T$ is the temperature and $V$ is the volume of the source.
For a source expanding spherically, it is assumed that $a=C_1R=C_1(R_0+\alpha t)$
\cite{LiuRuZhangWong-JPG14}, where $C_1$ is the source-size parameter, $R_0$ is the
initial source radius, and $\alpha$ is a parameter related to the average expansion
velocity of the source. With a hydrodynamical calculation for $R_0=6$~fm and
$T_0=170$~MeV, the model parameters $\gamma$ and $\alpha$ are fixed at 1.627 and
0.62 \cite{LiuRuZhangWong-JPG14}, respectively. In the model calculations of this
paper, the values of $C_1$ are taken to be 0.35 and 0.40 as in Refs. \cite{LiuRuZhangWong-JPG14,BaryRuZhang-JPG18}.

For the EPG source with Bose-Einstein condensation, the functions are
\cite{BaryRuZhang-JPG18}
\begin{equation}
\label{R12_1}
R_2(\mp_i,\mp_j)=\frac{|G^{(1)}(\mp_i,\mp_j)|^2-N_0^2|u_0(\mp_i)|^2|u_0(\mp_j)|^2}
{G^{(1)}(\mp_i,\mp_i)\,G^{(1)}(\mp_j,\mp_j)},
\end{equation}
\begin{eqnarray}
\label{R123_1}
&&R_3(\mp_i,\mp_j,\mp_k)=\nonumber\\
&&\hspace*{5mm}2\frac{{\rm Re}\big[G^{(1)}(\mp_i,\mp_j)G^{(1)}(\mp_j,\mp_k)G^{(1)}(\mp_k,
\mp_i) -N_0^3 f_3(\mp_i,\mp_j,\mp_k)\big]}{G^{(1)}(\mp_i,\mp_i)G^{(1)}(\mp_j,\mp_j)
G^{(1)}(\mp_k,\mp_k)},~~~~
\end{eqnarray}
and
\begin{eqnarray}
\label{R1234_1}
&&R_4(\mp_i,\mp_j,\mp_k,\mp_l)=\nonumber\\
&&\hspace*{5mm}2\frac{{\rm Re}\big[G^{(1)}\!(\mp_{\!i},\mp_{\!j})G^{(1)}\!(\mp_{\!j},
\mp_{\!k})G^{(1)}\!(\mp_{\!k},\mp_{\!l})G^{(1)}\!(\mp_{\!l},\mp_{\!i})
-N_0^4 f_4(\mp_{\!i},\mp_{\!j},\mp_{\!k},\mp_{\!l})\big]}{G^{(1)}(\mp_i,\mp_i)
G^{(1)}(\mp_j,\mp_j)G^{(1)}(\mp_k,\mp_k) G^{(1)}(\mp_l,\mp_l)},
\end{eqnarray}
where $G^{(1)}(\mp_i,\mp_j)$ is the one-particle density matrix, $N_0$ is the
ground-state particle number, and $u_n(\mp)~(n=0,1,2,\cdots)$ is the single-particle
wave function, and
\begin{equation}
\label{Gp1_1}
G^{(1)}(\mp_i,\mp_j)=\sum_n u_n^*(\mp_i) u_n(\mp_j)\frac{g_n \mathcal Z\,e^{-\tilde E_n/T}}
{1-\mathcal Z\,e^{-\tilde E_n/T}},
\end{equation}
where $g_n$ is the degeneracy, $\mathcal Z$ is the fugacity parameter including
the factor for the lowest energy level $\varepsilon_0$, and $\tilde E_n$ is the
eigenenergy of a relativistic pion relative to $\varepsilon_0$
\cite{NarGla-PRA99,WongZhang-PRC07,LiuRuZhangWong-JPG14,BaryRuZhang-JPG18}.
In Eqs.~(\ref{R123_1}) and (\ref{R1234_1}), $f_3(\mp_i,\mp_j,\mp_k)$ and $f_4(\mp_i,
\mp_j,\mp_k,\mp_l)$ are functions of $u_0(\mp)$, $G^{(1)}(\mp_i,\mp_j)$, and $N_0$
\cite{BaryRuZhang-JPG18}.
For a completely chaotic source, $N_0<<N$, the second terms in the numerators in
Eqs.~(\ref{R12_1}), (\ref{R123_1}), and (\ref{R1234_1}) approach 0.  However, in the
nearly completely coherent case, almost all particles are in the ground condensate
state, functions $f_3(\mp_i,\mp_j,\mp_k)\to |u_0(\mp_i)|^2|u_0(\mp_j)|^2|u_0(\mp_k)|^2$
and $f_4(\mp_i,\mp_j,\mp_k,\mp_l) \to |u_0(\mp_i)|^2 |u_0(\mp_j)|^2| u_0(\mp_k)|^2
|u_0(\mp_l)|^2$, and the two terms in the numerators in Eqs.~(\ref{R12_1}), (\ref{R123_1}), and (\ref{R1234_1}) approximately cancel each other.
Therefore, the two-pion, three-pion, and four-pion correlation functions approach
1 in the completely coherent case \cite{LiuRuZhangWong-JPG14}.
From Eq.~(\ref{Gp1_1}), we can calculate the density matrices
$G^{(1)}(\mp_i,\mp_j)$ for the EPG source at each evolution step (with given
temperature $T$ and total particle number $N$) with the technique developed in
Ref. \cite{WongZhang-PRC07}, and then obtain the two-, three-, and four-pion
correlation functions \cite{WongZhang-PRC07,LiuRuZhangWong-JPG14,BaryRuZhang-JPG18}.

The normalized three-pion correlation function $r_3(\mp_1,\mp_2,\mp_3)$ is defined
by dividing $R_3(\mp_1,\mp_2,\mp_3)$ by the square root of the product of the
two-particle correlators \cite{HeiZhaSug}:
\begin{equation}
\label{smallr3}
r_3(\mp_1,\mp_2,\mp_3)=\frac{R_3(\mp_1,\mp_2,\mp_3)}{\sqrt{R_2(\mp_1,\mp_2)R_2(\mp_2,
\mp_3)R_2(\mp_3,\mp_1)}}.
\end{equation}
Function $r_3$ is insensitive to resonance decay
\cite{HeiZhaSug,NA44,WA98,STAR-PRL03,ALICE-PRC14}, and is directly related to the
condensation fraction for our space-symmetric EPG sources. Similarly, the normalized
four-pion correlation function $r_4(\mp_1,\mp_2,\mp_3,\mp_4)$ is defined by
\cite{Gangadharan15}
\begin{equation}
\label{smallr4}
r_4(\mp_1,\mp_2,\mp_3,\mp_4)=\frac{R_{44}(\mp_1,\mp_2,\mp_3,\mp_4)}{\sqrt{R_2(\mp_1,
\mp_2)R_(\mp_2,\mp_3)R_2(\mp_3,\mp_4)R_2(\mp_4,\mp_1)}},
\end{equation}
where
\begin{equation}
R_{44}(\mp_1,\mp_2,\mp_3,\mp_4)=R_4(\mp_1,\mp_2,\mp_3,\mp_4)+R_4(\mp_1,\mp_2,\mp_4,
\mp_3)+R_4(\mp_1,\mp_3,\mp_2,\mp_4).
\end{equation}

In the EPG model, the intercept of $R_2(\mp_1,\mp_2)$ at zero relative momentum can be
written as
\cite{LiuRuZhangWong-JPG14}
\begin{equation}
\lambda(\mp)=R_2(\mp,\mp)=1-\frac{N_0^2|u_0(\mp)|^4}{G^{(1)}(\mp,\mp)^2} \equiv
1-[f_0 F_N(\mp)]^2,
\label{lamFN}
\end{equation}
where $f_0=N_0/N$ is the condensation fraction and
\begin{equation}
F_N(\mp)=N|u_0(\mp)|^2/G^{(1)}(\mp,\mp).
\label{FN}
\end{equation}
Hence, the intercepts of $r_3$ and $r_4$ at zero relative momentum
$q_{ij}=\sqrt{-(p_i-p_j)^\mu(p_i-p_j)_\mu}=0,\,(i,j=1,2,3,4)$ can be written as
\begin{equation}
I_3(\mp)\equiv r_3(\mp,\mp,\mp)=2\,\frac{1\!-\!3[f_0 F_N(\mp)]^2+2[f_0 F_N(\mp)]^3}
{\big[1-[f_0 F_N(\mp)]^2\big]^{3/2}}
\end{equation}
and
\begin{equation}
I_4(\mp)\equiv r_4(\mp,\mp,\mp,\mp)=6\,\frac{1\!-\!6[f_0 F_N(\mp)]^2+8[f_0 F_N(\mp)]^3
-3[f_0 F_N(\mp)]^4}{\big[1-[f_0 F_N(\mp)]^2\big]^{2}}.
\end{equation}
These intercepts are functions of condensation fraction $f_0$ and particle momentum
$\mp$, and thus are functions of system temperature $T$, particle number $N$,
source-size parameter $C_1$ \cite{LiuRuZhangWong-JPG14,BaryRuZhang-JPG18}, and particle
momentum $\mp$.

\begin{figure}[htb]
\includegraphics[width=0.85\columnwidth]{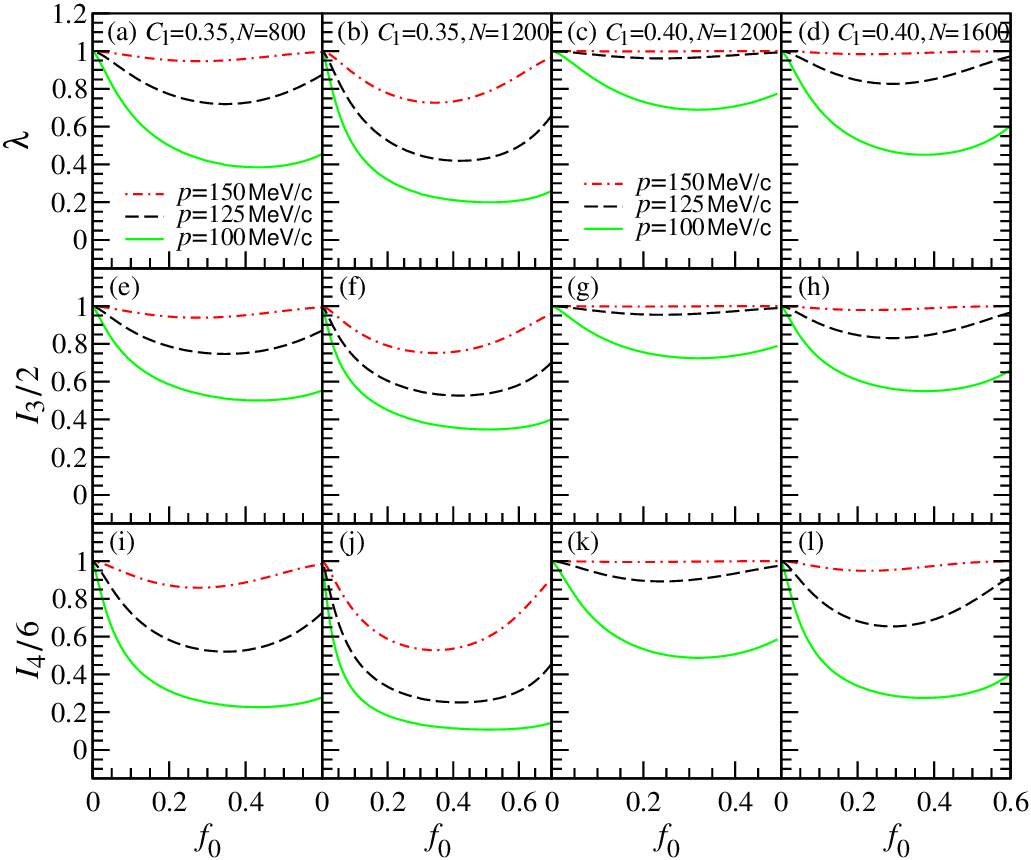}
\caption{(Color online) Intercepts of two-pion correlation function, $\lambda$, and
of normalized three- and four-pion correlation functions, $I_3$ and $I_4$,
as functions of condensation fraction $f_0$ for EPG sources with different values
of source-size parameter $C_1$, particle number $N$, and particle momentum $p$. }
\label{fI234-f0}
\end{figure}

\begin{figure}[htb]
\includegraphics[width=0.85\columnwidth]{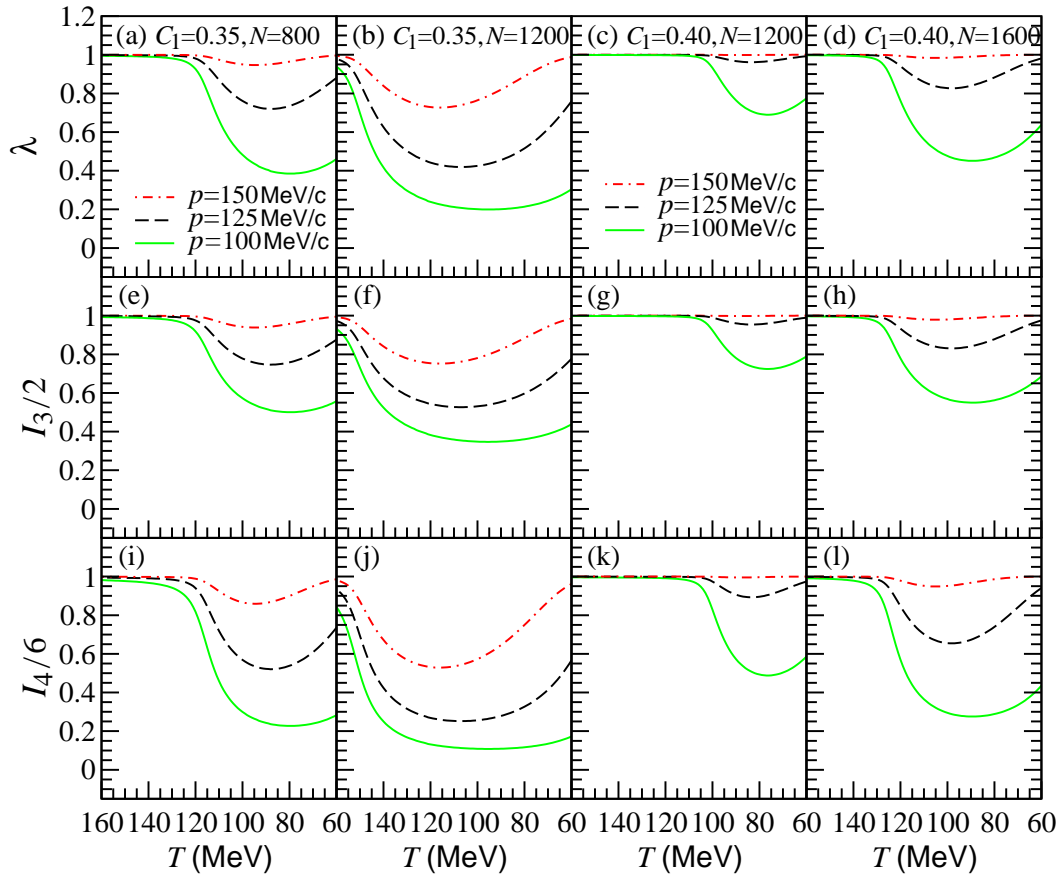}
\caption{(Color online) Intercepts of two-pion correlation function, $\lambda$, and
of normalized three- and four-pion correlation functions, $I_3$ and $I_4$,
as functions of temperature $T$ for EPG sources with different values of source-size
parameter $C_1$, particle number $N$, and particle momentum $p$. }
\label{fI234-T}
\end{figure}

We plot in Figs.~\ref{fI234-f0}(a)--(d), \ref{fI234-f0}(e)--(h), and \ref{fI234-f0}(i)--(l) the intercepts $\lambda$, $I_3$, and $I_4$, respectively,
as functions of the condensation fraction $f_0$
for EPG sources with different values of source-size parameter $C_1$,
particle number $N$, and particle momentum $p=|\mp|$.  The variational
tendencies of $\lambda(f_0)$, $I_3(f_0)/2$, and $I_4(f_0)/6$ are almost the same.
They are 1 when $f_0=0$.  As $f_0$ increases from 0, the intercepts decrease to
their minima, and then increase with increasing $f_0$.  The decreases of the
intercepts are much smaller for higher momentum.  This is because the particles
with higher momenta are likely emitted from the excited states incoherently,
even from a source with finite $f_0$. For the same $f_0$ value, the intercepts
are smaller for the higher particle numbers and the smaller source-size parameter.
This is due to the function $F_N$ defined in Eq.~(\ref{FN}), which increases with
decreasing $N$ and increasing $C_1$ at low momentum in the EPG model (see Fig.~7
in Ref. \cite{LiuRuZhangWong-JPG14}). From Eq.~(\ref{lamFN}) we see that the
intercept increases with increasing $f_0$ if $F_N(p)$ deceases more rapidly with
increasing $f_0$.  This is the reason for the increases of the intercepts for
higher momentum at high $f_0$ (with low source temperature).

We further plot in Figs.~\ref{fI234-T}(a)--(d), \ref{fI234-T}(e)--(h), and \ref{fI234-T}(i)--(l) the intercepts $\lambda$, $I_3$, and $I_4$, respectively,
as functions of the source temperature $T$ for EPG sources with different values
of $C_1$, $N$, $p$.  Because $T$ and $f_0$ have an antilinear relationship (see
Fig.~19 in Ref. \cite{BaryRuZhang-JPG18}), the intercepts $\lambda(T)$, $I_3(T)/2$,
and $I_4(T)/6$ have similar variations with decreasing $T$ to those
in Fig.~\ref{fI234-f0} with increasing $f_0$.  They are 1 at high temperature and
decrease to their minima at low temperature.  The minima decrease with decreasing
$p$, increasing $N$, and decreasing $C_1$.

\section{Results of normalized MPC functions}
In this section we analyze the normalized MPC functions $r_3(Q_3)$ and $r_4(Q_4)$ in
different regions of the average transverse momenta $K_{T3}$ and $K_{T4}$ in the EPG
model, and compare the model results with corresponding experimental data
\cite{ALICE-PRC14}.
Here,
\begin{equation}
Q_3=\sqrt{q_{12}^2+q_{13}^2+q_{23}^2},
\end{equation}
\begin{equation}
Q_4=\sqrt{q_{12}^2+q_{13}^2+q_{14}^2+q_{23}^2+q_{24}^2+q_{34}^2},
\end{equation}
\begin{equation}
K_{T3}=\frac{|\mp_{1T}+\mp_{2T}+\mp_{3T}|}{3},
\end{equation}
\begin{equation}
K_{T4}=\frac{|\mp_{1T}+\mp_{2T}+\mp_{3T}+\mp_{4T}|}{4}.
\end{equation}

\subsection{Results for $r_3(Q_3)$}

\begin{figure}[htb]
\includegraphics[width=0.7\columnwidth]{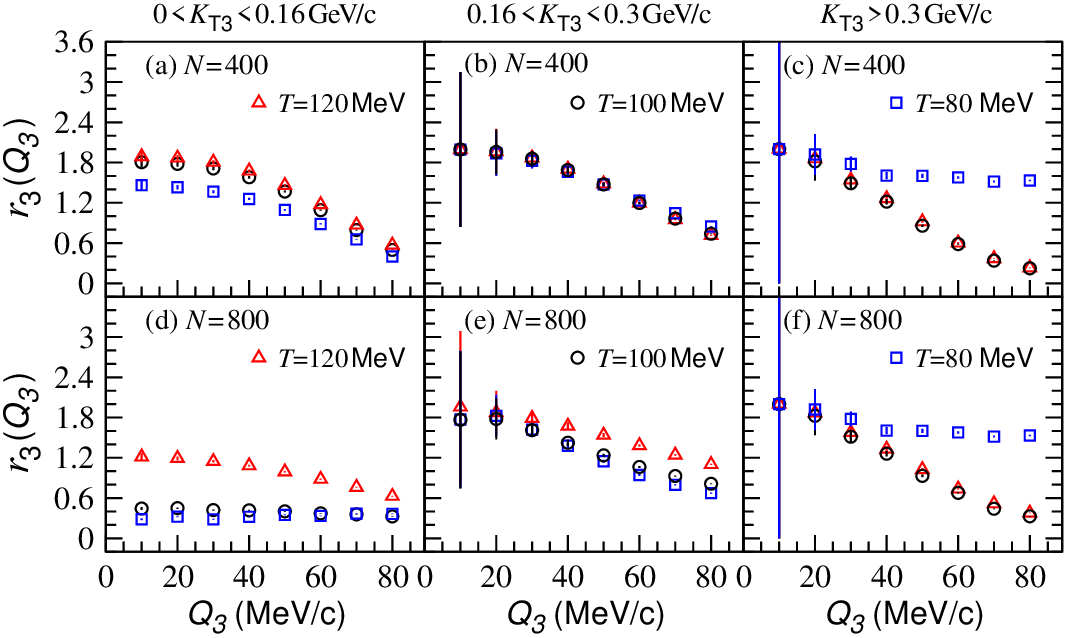}
\caption{(Color online) Normalized three-pion correlation function $r_3(Q_3)$
for the EPG sources with different temperatures and particle numbers and in the
transverse-momentum regions $0<K_{T3}<0.16$~GeV/$c$, $0.16<K_{T3}<0.3$~GeV/$c$,
and $K_{T3}>0.3$~GeV/$c$. Here, the source-size parameter is $C_1=0.35$. }
\label{fr335}
\end{figure}

We plot in Fig.~\ref{fr335} the normalized three-pion correlation function $r_3(Q_3)$
for different source temperatures and small and large particle numbers, $N=$~400 and
800, in the EPG model with $C_1=0.35$. In the low average-transverse-momentum region
$0<K_{T3}<0.16$~GeV/$c$, $r_3(Q_3)$ decreases with decreasing $T$ and is lower for
high $N$.  This is because the system has more condensation at lower temperature
and higher particle number than at higher temperature and lower particle number.
For the low particle number, $r_3(Q_3)$ decreases with increasing $Q_3$.  However, for
high particle number and low temperature, $T=80$~MeV, $r_3(Q_3)$ increases slightly
with increasing $Q_3$.  In the intermediate average-transverse-momentum region
$0.16<K_{T3}<0.3$~GeV/$c$, $r_3(Q_3)$ results are higher than those in the low
average-transverse-momentum region, and the dependences of $r_3(Q_3)$ on the source
temperature and particle number become weaker than those in the low
average-transverse-momentum region. These reflect the important characteristic
of the EPG source that the particles with high momenta are likely emitted from
excited states thermally and incoherently even for the source with a considerable
condensation fraction. In the high average-transverse-momentum region $K_{T3}>
0.3$~GeV/$c$, $r_3(Q_3)$ is almost the same for the temperatures $T=$~120 and
100~MeV.  However, $r_3(Q_3)$ becomes flat at large $Q_3$ for the low temperature
$T=80$~MeV.

To explain the variational tendency of $r_3(Q_3)$ with increasing $Q_3$, we consider
the special case $\mp_1=\mp_2$ in Eq.~(\ref{smallr3}) for simplicity.  In this case,
we have
\begin{eqnarray}
r_3(Q_3)&=&\frac{2}{\sqrt{\lambda(p_2)}}\left[1-\frac{2[1-\lambda(p_2)]^{3/4}
[1-\lambda(p_3)]^{1/4}}{\sqrt{R_2^{ch}(q_{23})}+[1-\lambda(p_2)]^{1/4}
[1-\lambda(p_3)]^{1/4}}\right]\cr
&\approx &\frac{2}{\sqrt{\lambda(\bar p)}}\left[1-\frac{2[1-\lambda(\bar p)]}
{\sqrt{R_2^{ch}(q_{23})}+\sqrt{1-\lambda(\bar p)}}\right]\cr
&=&\frac{2}{\sqrt{\lambda(\bar p)}}\left[1-\frac{2\sqrt{1-\lambda(\bar p)}}
{\sqrt{R_2^{ch}(q_{23})}/\sqrt{1-\lambda(\bar p)}+1}\right],
~~~~(Q_3=\sqrt{2}\,q_{23}),
\label{r3spe}
\end{eqnarray}
where $0<R_2^{ch}(q_{23})<1$ is the two-pion correlator of completely chaotic
source, and $\lambda(\bar p)$ is the chaoticity parameter (intercept) of the
two-pion correlation at average particle momentum $\bar p$.  For a completely
chaotic source, $f_0=0$, $\lambda(\bar p)=1$, and $r_3(Q_3)=2$.  For finite $f_0$
and fixed $\lambda(\bar p)$, $r_3(Q_3)$ decreases with increasing $Q_3$ because
$R_2^{ch}(q_{23})$, as a function of source size and $Q_3$, decreases with
increasing $Q_3$.  In fact, the value of $\lambda$ in Eq.~(\ref{r3spe}) is
$Q_3$-dependent because $\bar p$ is related to $Q_3$.  The average particle
momentum $\bar p$ will increase with increasing $Q_3$ if there are no other
constraints. This leads to an increasing $\lambda(\bar p)$ (see Fig.~\ref{fI234-f0})
and decreasing $[1-\lambda(\bar p)]$ with increasing $Q_3$. From Eq.~(\ref{r3spe})
we see that $r_3(Q_3)$ will increase with increasing $Q_3$ if $[1-\lambda(\bar p)]$ decreases with increasing $Q_3$ faster than $R_2^{ch}(q_{23})$ does.  This may
occur at low temperature, where $\lambda(\bar p)$ decreases rapidly with increasing
particle momentum (see Fig.~\ref{fI234-T}).

\begin{figure}[htb]
\includegraphics[width=0.7\columnwidth]{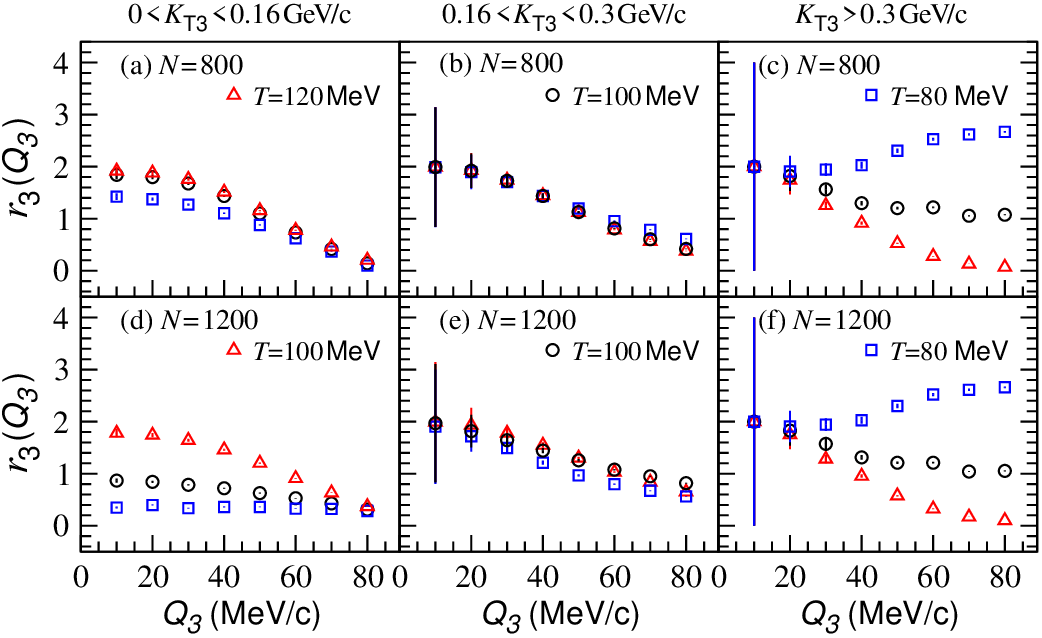}
\caption{(Color online) Normalized three-pion correlation function $r_3(Q_3)$
for the EPG sources with different temperatures and particle numbers and in the
transverse-momentum regions $0<K_{T3}<0.16$~GeV/$c$, $0.16<K_{T3}<0.3$~GeV/$c$,
and $K_{T3}>0.3$~GeV/$c$. Here, the source-size parameter is $C_1=0.40$.  }
\label{fr340}
\end{figure}

We plot in Fig.~\ref{fr340} the normalized three-pion correlation function $r_3(Q_3)$
for EPG sources with $C_1=0.40$ and $N=$~800 and 1200.  The behaviors of $r_3(Q_3)$
in the low and intermediate average-transverse-momentum regions are similar to those
in Fig.~\ref{fr335}.  In the high average-transverse-momentum region, the results of
$r_3(Q_3)$ for the low temperatures obviously increase with increasing $Q_3$ at large
$Q_3$ compared to the results for the high temperature. This is related to the
increase of $\lambda$ with increasing particle momentum in the wide momentum
variational region.

\begin{figure}[htb]
\vspace*{5mm}
\includegraphics[width=0.7\columnwidth]{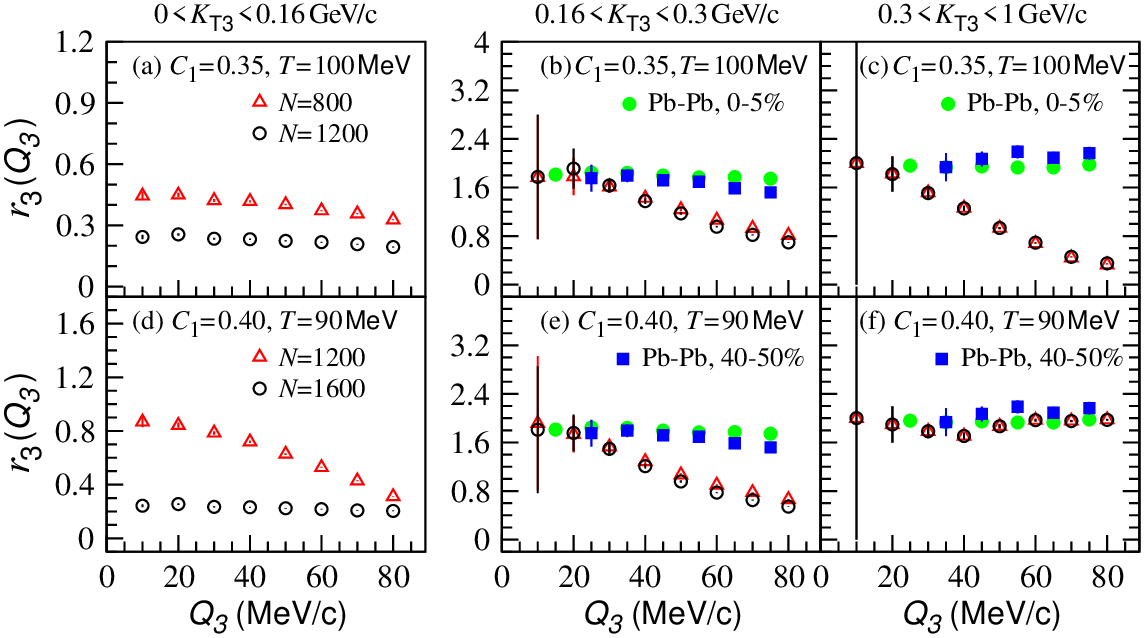}
\caption{(Color online) Top panels: Normalized three-pion correlation function
$r_3(Q_3)$ for the EPG sources with $C_1=0.35$, $T=100$~MeV, and $N=$~800 and
1200. Bottom panels: Normalized three-pion correlation function $r_3(Q_3)$ for
the EPG sources with $C_1=0.40$, $T=90$~MeV, and $N=$~1200 and 1600. The solid
circle and square symbols in the middle and right panels are for the data in
the central (0-5\%) and peripheral (40-50\%) Pb-Pb collisions at the LHC
\cite{ALICE-PRC14}. }
\label{fr3e}
\end{figure}

We plot in Fig.~\ref{fr3e} the normalized three-pion correlation function $r_3(Q_3)$
for EPG sources in the average-transverse-momentum regions $0<K_{T3}<0.16$~GeV/$c$,
$0.16<K_{T3}<0.3$~GeV/$c$, and $0.3<K_{T3}<1$~GeV/$c$.  Here, the values of
temperature $T$ and particle number $N$ for the source with parameters $C_1=$~0.35
and 0.40 are taken as the same in Ref.~\cite{BaryRuZhang-JPG18} where the model
results of MPCs $C_3(Q_3)$, $c_3(Q_3)$, $C_4(Q_4)$, $a_4(Q_4)$, $b_4(Q_4)$, and
$c_4(Q_4)$ are compared with the experimental data in the average-transverse-momentum
regions $0.16<K_{T3}<0.3$~GeV/$c$ and $0.3<K_{T3}<1$~GeV/$c$ \cite{ALICE-PRC16}.
In the low transverse-momentum region (Figs.~\ref{fr3e}(a) and (d)), the results
for the higher particle numbers are lower than those for the lower particle numbers
because of severe condensation for the sources with higher $N$. However, in the intermediate and high transverse-momentum regions, the differences of $r_3(Q_3)$
results for the lower and higher $N$ values are small (see Figs.~\ref{fr3e}(b) and
(e)) and the results are almost the same (see Figs.~\ref{fr3e}(c) and (f)).
This is because the particles with high momenta are likely emitted from the excited
states incoherently.
In Fig.~\ref{fr3e}(b), (c), (e), and (f), the solid circles and squares denote the
experimental $r_3$ data for central and peripheral Pb-Pb collisions, respectively,
at the LHC \cite{ALICE-PRC14}.  The experimental results are almost independent of
collision centrality and almost flat with increasing $Q_3$.  At small $Q_3$, the
results of the EPG model agree with the experimental data.  Furthermore, the model results in Fig.~\ref{fr3e}(f) almost reproduce the experimental data in the high transverse-momentum region.  As discussed above, the variational tendency of
$r_3(Q_3)$ is related to the source size and $\lambda$ increase with particle
momentum.  Because the EPG model considers only a simple source expanding spherically,
it is unpractical to hope the model results can completely reproduce the experimental data.

\subsection{$r_4(Q_4)$ results}

\begin{figure}[htb]
\includegraphics[width=0.7\columnwidth]{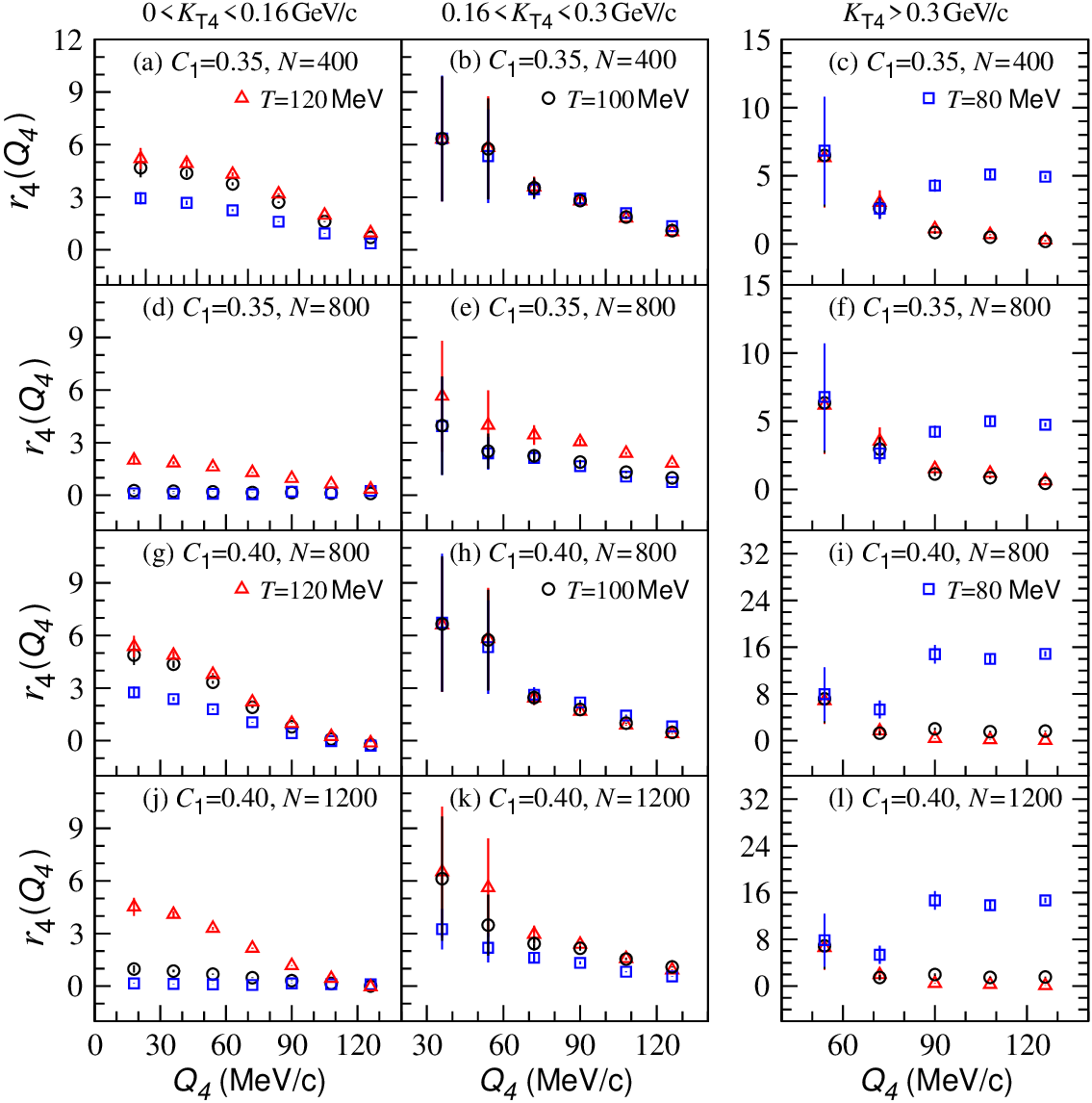}
\caption{(Color online) Normalized four-pion correlation functions $r_4(Q_4)$ for
the EPG sources with different source temperatures and particle numbers, and in
the transverse-momentum intervals $0<K_{T4}<0.16$~GeV/$c$, $0.16<K_{T4}<0.3$~GeV/$c$,
and $K_{T4}>0.3$~GeV/$c$. The source-size parameters $C1$ are 0.35 (the first and
second rows) and 0.40 (the third and fourth rows). }
\label{fr43540}
\end{figure}

We plot in Fig.~\ref{fr43540} the normalized four-pion correlation function $r_4(Q_4)$
for EPG sources with different temperatures and particle numbers as in Figs.~\ref{fr335} and \ref{fr340}.  The variation of $r_4(Q_4)$ as a function of $Q_4$ is similar to that
of $r_3(Q_3)$ as a function of $Q_3$. In the low average-transverse-momentum region
$0<K_{T4}<0.16$~GeV/$c$, $r_4(Q_4)$ decreases with decreasing temperature.
In the intermediate average-transverse-momentum region $0.16<K_{T4}<0.3$~GeV/$c$, $r_4(Q_4)$ decreases with decreasing temperature for lower particle number.
However, the results for the high particle numbers are almost independent of
temperature.  In the high average-transverse-momentum region $K_{T4}>0.3$~GeV/$c$,
$r_4(Q_4)$ is obviously enhanced at large $Q_4$ for the low temperature $T=80$~MeV.
This is because the chaoticity parameter of two-pion HBT correlations, $\lambda$,
increases rapidly with increasing particle momentum at large $Q_4$ in the EPG model.

\begin{figure}[htb]
\includegraphics[width=0.65\columnwidth]{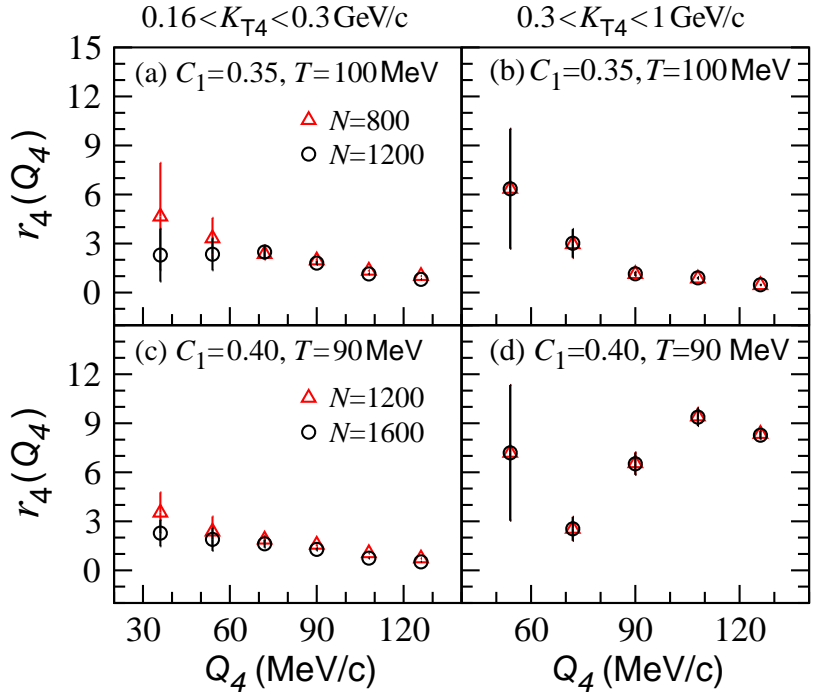}
\caption{(Color online) Normalized four-pion correlation functions $r_4(Q_4)$ for
the EPG sources with different particle numbers and in the transverse-momentum
intervals $0.16<K_{T4}<0.3$~GeV/$c$ and $0.3<K_{T4}<1$~GeV/$c$. The source
temperature are 100 and 90~MeV when $C1=$~0.35 and 0.40, respectively. }
\label{fr4e}
\end{figure}

In figure \ref{fr4e} we plot the normalized four-pion correlation function $r_4(Q_4)$
for EPG sources in the average-transverse-momentum regions $0.16<K_{T4}<0.3$~GeV/$c$
and $0.3<K_{T4}<1$~GeV/$c$.  Here, we use the source temperatures and particle numbers
as in Ref.~\cite{BaryRuZhang-JPG18} in comparing the MPC model results with
the experimental Pb-Pb collision data \cite{ALICE-PRC16}.
In the low transverse-momentum region, the results of $r_4(Q_4)$ for the large particle
numbers are lower than 6 at small $Q_4$.  This indicates that there are considerable
condensations for the EPG sources with the high $N$. In the high transverse-momentum
region, $r_4(Q_4)$ is almost independent of particle number because of the
characteristic of the EPG sources that the particles with high momenta are likely
emitted from the excited states incoherently even for the source with a considerable
condensation fraction.  For the source with $C_1=0.40$ and $T=90$~MeV, $r_4(Q_4)$ is
significantly enhanced at large $Q_4$, which reflects the two-particle correlation decreases with increasing particle momentum in the EPG model.
Because the model results of $r_3(Q_3)$ in the high transverse-momentum region almost
reproduces the experimental data \cite{ALICE-PRC16}, we hope the enhancement of $r_4(Q_4)$ in this transverse-momentum region will be identified in future experimental
data analyses.

\section{Summary and discussions}
Pion multiplicity have been observed to reach several thousand in heavy-ion collisions
at the LHC.  This high pion multiplicity possibly causes significant system condensation and leads to a partially coherent pion-emitting source. The normalized MPC functions
$r_3(Q_3)$ and $r_4(Q_4)$ are useful for exploring the coherence of pion-emitting
sources produced in high-energy heavy-ion collisions. On the basis of our previous
MPC analyses in the EPG model with Bose-Einstein condensation in relativistic heavy-ion
collisions, we have investigated the normalized three- and four-pion correlation
functions in different average-transverse-momentum regions of three and four particles, and studied the effects of the source temperature $T$ and particle number $N$ on the
normalized MPC functions in the EPG model. We have found that the intercepts of the
normalized MPC functions at $Q_{3,4}=0$ are related to the chaoticity parameter of
two-pion correlation, $\lambda$, and sensitive to the condensation of the EPG source.
The values of the normalized MPC functions in low average-transverse-momentum region decrease with decreasing $T$ and increasing $N$, because the source condensation
increases with decreasing $T$ and increasing $N$. However, these dependences of the normalized MPC functions on the source temperature and particle number become weak
in an intermediate average-transverse-momentum region, which reflects the important
characteristic of the EPG source that the particles with high momenta are likely
emitted from the excited states incoherently even for the source with a considerable
condensation fraction. In high average-transverse-momentum region, the normalized MPC
functions for low source temperatures are enhanced at larger relative momenta because
of the rapid increase of the two-pion chaoticity parameter $\lambda$ with increasing particle momentum in the EPG model.

Finally, let us make an estimation of the average phase-space density $\langle f
\rangle_p$ for the EPG sources with different values of $C_1$ parameter, particle
number $N$, and source temperature $T$. Based on the method proposed by G.~F.~Bertsch
\cite{Bertsch9496}, the phase-space density can be estimated with single-particle
momentum distribution and two-particle HBT radium. In the EPG model, it is assumed
that the relaxation time of the system is smaller than the source evolution time and
the expansion of the pion gas may approximately deal with a quasi-static process
\cite{LiuRuZhangWong-JPG14,BaryRuZhang-JPG18}. Therefore, at each time during the
source evolution, there is a certain system temperature $T$ (see the Fig.~1 in Ref.
\cite{LiuRuZhangWong-JPG14}), and the corresponding single-pion momentum distribution
is $(d^3N/d^3p)(T)\!=\!G^{(1)}(\mp,\mp)$ \cite{WongZhang-PRC07,LiuRuZhangWong-JPG14,BaryRuZhang-JPG18}.
Using the parameterized two-pion correlation function, $1\!+\!\lambda \exp(-q^2
R_{\rm HBT}^2)$, for the spherical EPG sources, the average phase-space density
is given by \cite{Bertsch9496}, $\langle f\rangle_p\!=\!(d^3N/d^3p)\,\lambda\,
\sqrt{\pi}/(4R_{\rm HBT}^3)$. We can calculate $(d^3N/d^3p)(T)$, $\lambda(p,T)$,
and $R_{\rm HBT}(p,T)$ \cite{WongZhang-PRC07,LiuRuZhangWong-JPG14,BaryRuZhang-JPG18}
for the EPG sources, and obtain $\langle f\rangle_{p=100 {\rm MeV\!/c}}=1.509$ and
$\langle f\rangle_{p=500 {\rm MeV\!/c}}=0.006$ for the parameter set ($C_1=0.40$,
$N=1200$, and $T=100$~MeV/c). The average phase-space density decreases greatly
with increasing particle momentum. The high $\langle f\rangle_p$ at small particle
momentum corresponds to a condensation. The average phase-space density at the
average momentum, $\langle f\rangle_{\langle p\rangle}$, for the low and high
source temperatures $T=$~100 and 150~MeV are 0.030 and 0.016 for the parameter
set ($C_1=0.40$ and $N=1200$); 0.060 and 0.022 for the parameter set ($C_1=0.40$
and $N=1600$); and 0.108 and 0.028 for the parameter set ($C_1=0.35$ and $N=1200$).
The average phase-space density decreases with increasing temperature and decreasing
particle number. The values of $\langle f\rangle_{\langle p\rangle}$ for the smaller
source-size parameter $C_1=0.35$ are higher than those for the larger source-size
parameter $C_1=0.40$.

In relativistic heavy-ion collisions, coherent emission may arise from the
formation of a disoriented chiral condensate (DCC) \cite{GreGonMul-PLB93,Bjorken-APPB97,Rajagopal-11}, pionic or gluonic Bose-Einstein
condensations \cite{{WongZhang-PRC07,LiuRuZhangWong-JPG14,BaryRuZhang-JPG18,
Begun14-15,Blaizot-NPA12}}, or multiple coherent sources from pulsed radiation \cite{Ikonen-PRC08}. Our previous investigations \cite{BaryRuZhang-JPG18} indicate
that the EPG model with pion condensation can approximately reproduced the MPCs in
Pb-Pb collisions at the LHC \cite{ALICE-PRC16}. In this study, we found that the
EPG model gives intercepts of the normalized MPC functions in agreement with the
experimental Pb-Pb collision data \cite{ALICE-PRC14}. The function $r_3(Q_3)$ in
the EPG model also approximately reproduces the experimental data in the high average-transverse-momentum region \cite{ALICE-PRC14}. These EPG-model results
indicate that the simple spherical EPG model may catch hold of some main
characteristics of the pion-emitting sources and the system produced in heavy-ion
collisions at the LHC may have a considerable condensation. As a result of the EPG model, we hope the significant enhancement of the normalized four-pion correlation
function $r_4(Q_4)$ at large relative momentum will be identified experimentally
in future. On the other hand, viscous hydrodynamics has widely been used to describe
the system evolution in relativistic heavy-ion collisions. It will be of interest
to develop a model of identical pion-emitting source that evolves with viscous
hydrodynamics.

\begin{acknowledgments}
This research was supported by the National Natural Science Foundation of China under
Grant Nos. 11675034 and 11275037, and the China Scholarship Council.
Mark Kurban, M. Sc., from Liwen Bianji, Edanz Editing China (www.liwenbianji.cn/ac),
edited a draft of this manuscript.
\end{acknowledgments}

\end{document}